\documentclass[9pt,conference]{IEEEtran}
\usepackage{amssymb,amsthm,amsmath,array}
\usepackage{graphicx}
\usepackage{xspace}
\usepackage[sort&compress, numbers]{natbib}
\usepackage{stmaryrd}
\usepackage{xcolor}
\usepackage{mathtools}
\usepackage{float}
\usepackage{textcomp}

\usepackage{amsmath}

\usepackage{amsthm}
\usepackage{amssymb}
\usepackage{amsfonts}
\usepackage{dsfont}
\usepackage{accents}
\usepackage{bm}
\usepackage{hyperref}

\usepackage{tabularx} 
\usepackage{caption}
\usepackage{multirow}
\usepackage{booktabs}
\usepackage{graphicx}
\graphicspath{{Figures/}}
 \usepackage{subcaption}

 \usepackage{tcolorbox}
 \usepackage{wasysym}
  \usepackage{longtable}

\usepackage{tikz,pgfplots}
\usepgfplotslibrary{groupplots}
\usepgfplotslibrary{fillbetween}
\usetikzlibrary{intersections,calc,arrows,matrix,spy}
\usepackage{color}
\definecolor{darkred}{rgb}{0.6,0,0}
\definecolor{darkgreen}{rgb}{0,0.5,0}
\definecolor{darkblue}{rgb}{0,0,0.5}
\definecolor{SkyBlue}{rgb}{0.53, 0.81, 0.92}
\pgfplotsset{compat=1.5.1}

\newcommand\notsotiny{\@setfontsize\notsotiny\@vipt\@viipt}
\pgfplotsset{every tick label/.append style={font=\notsotiny}}
\tikzset{fontscale/.style = {font=\notsotiny}
    }

\usepackage{algorithm}
\usepackage{algpseudocode}
\usepackage[squaren,Gray]{SIunits}

\newcommand{\vy}{\ensuremath{\boldsymbol{y}}}
\newcommand{\vr}{\ensuremath{\boldsymbol{r}}}
\newcommand{\vp}{\ensuremath{\boldsymbol{p}}}

\newcommand{\vR}{\ensuremath{\boldsymbol{R}}}

\usetikzlibrary{plotmarks}
\usepgflibrary{plotmarks}
\usepackage[normalem]{ulem}

\def\FBPColor{red}
\def\methodColor{blue}

\def\FBPmark{o}
\def\methodmark{square}

\begin{document}
\title{Localized Supervised Learning for  Cryo-ET Reconstruction}
\author{\IEEEauthorblockN{
       Vinith Kishore\IEEEauthorrefmark{1}, 
        Valentin Debarnot\IEEEauthorrefmark{2}, 
        AmirEhsan Khorashadizadeh\IEEEauthorrefmark{1}
        and Ivan Dokmani\'c \IEEEauthorrefmark{1}
    }
    \IEEEauthorblockA{
        \IEEEauthorrefmark{1} Department of Mathematics and Computer Science, University of Basel, 4051 Basel, Switzerland.\\
        \IEEEauthorrefmark{2} CREATIS, INSA Lyon, University of Lyon, France.\\
}
}
\maketitle
\begin{abstract}
    Cryo-electron tomography (Cryo-ET) is a powerful tool in structural biology for 3D visualization of cells and biological systems at resolutions sufficient to identify individual proteins in situ. The measurements are collected by tilting the frozen specimen and exposing it to an electron beam of known dosage. As the biological samples are prone to electron damage, the samples can be exposed to only a limited dosage of electrons, leading to noisy and incomplete measurements. Thus, the reconstructions are noisy and incomplete, leading to the missing wedge problem. Currently, self-supervised learning is used to compensate for this issue. This typically involves, for each volume to recover, training a large 3D UNet on the initial noisy reconstruction, leading to large training time and memory requirements. In this work, we exploit the local nature of the forward model to train a lightweight network using only localized data from the measurements. This design provides flexibility in balancing computational and time requirements while reconstructing the volumes with high accuracy. We observe experimentally that this network can work well on unseen datasets, despite using a network trained on a few measurements.
\end{abstract}

\def\sz{4.5cm}
\def\ps{0.33}
\def\sc{268*0.5}
\begin{figure*}[h]
	\begin{subfigure}[t]{\ps\textwidth}
	    \centering
    	\begin{tikzpicture}[spy using outlines={circle,orange,magnification=4,size=2cm, connect spies}]
    		\node[ rotate=90] at (0,0) {}; outlines={circle,orange,magnification=2,size=3cm, connect spies}]
    		\node[rotate=0, line width=0.05mm, draw=white] at (0,0) { \includegraphics[height=\sz]{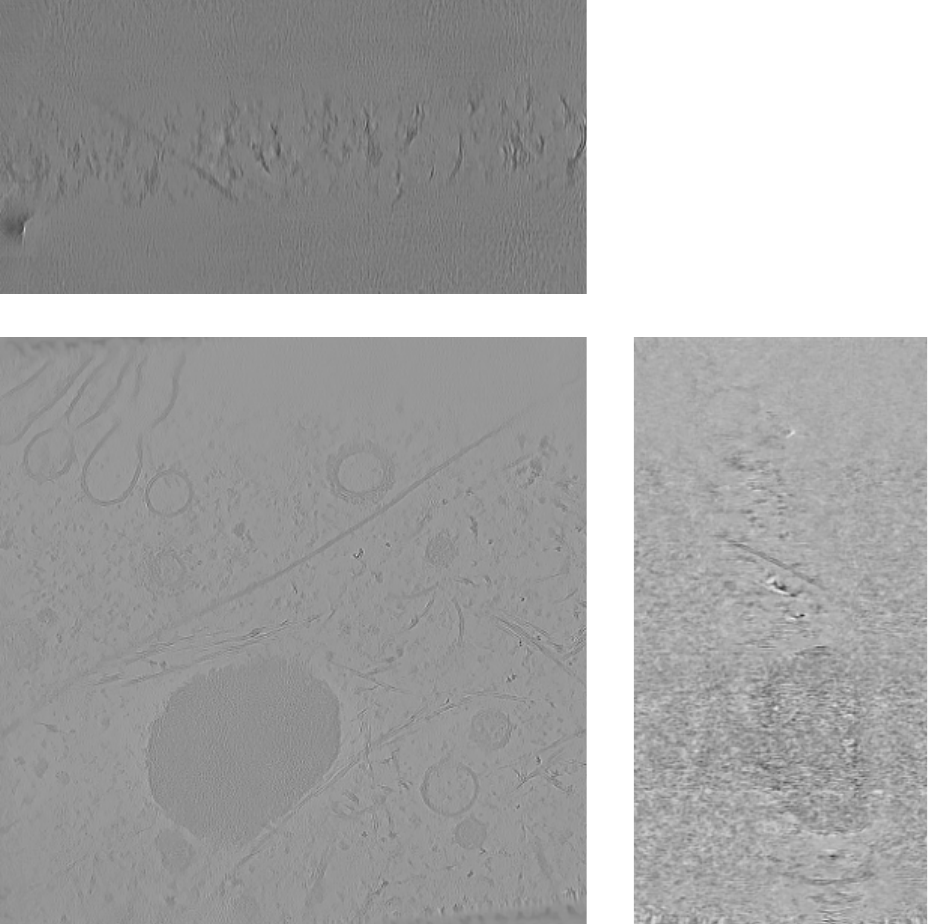}};
    		\spy on (-1.5,-0.2) in node [left] at (2.2,1.2);
		\end{tikzpicture} 
		\caption{Cryo-CARE+IsoNet (Reference)}
	\end{subfigure}\hfill
	\begin{subfigure}[t]{\ps\textwidth}
	    \centering
    	\begin{tikzpicture}[spy using outlines={circle,orange,magnification=4,size=2cm, connect spies}]
    		\node[ rotate=90] at (0,0) {}; outlines={circle,orange,magnification=2,size=3cm, connect spies}]
    		\node[rotate=0, line width=0.05mm, draw=white] at (0,0) { \includegraphics[height=\sz]{images/ref_test_data.pdf}};
    		\spy on (-1.5,-0.2) in node [left] at (2.2,1.2);
		\end{tikzpicture} 
		\caption{Ours}
	\end{subfigure}
        \begin{subfigure}[t]{\ps\textwidth}
	    \centering
          \begin{tikzpicture}
            \begin{axis}[
                width=0.99\linewidth, 
			grid=major, 
			grid style={dashed,gray!30}, 
			xlabel= resolution ($1/ \text{pixel size}$),
			ylabel=FSC,legend style={at={(0.5,0.4)}, legend cell align=right, align=right, draw=none,font=\scriptsize}]
            \addplot[mark=\FBPmark, mark size=1pt ,  mark repeat=20, color=\FBPColor] table [x expr=\coordindex/\sc, y=FBP, col sep=comma] {images/fsc_test_data.csv};
            \addlegendentry{FBP}
            \addplot[mark=\methodmark, mark size=1pt ,  mark repeat=20, color=\methodColor] table [x expr=\coordindex/\sc, y=OURS, col sep=comma] {images/fsc_test_data.csv};
            \addlegendentry{Ours}
            \end{axis}
            \end{tikzpicture}
        
		\caption{FSC Curve}
	\end{subfigure}\hfill
\caption{Evaluating our approach trained on real volume measurement pairs, on the test set of the real dataset. The FSC curves are computed using the volume recovered from Cryo-CARE + IsoNET as reference. From the FSC curve, we observe that the model can closely recover the reference reconstruction. }\label{fig:real-test}
\end{figure*}

\def\sz{3.8cm}
\def\ps{0.24}
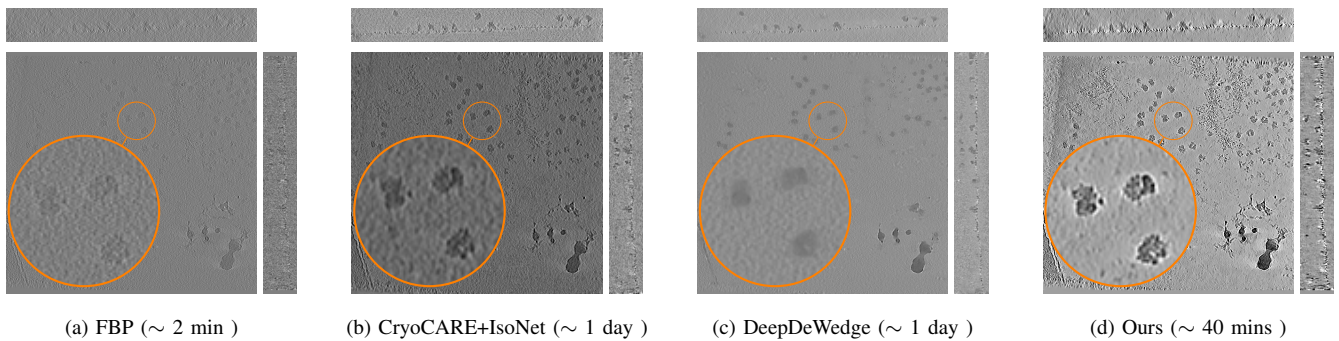
\begin{figure*}[h]
	\begin{subfigure}[t]{\ps\textwidth}
	    \centering
    	\begin{tikzpicture}[spy using outlines={circle,orange,magnification=4,size=2cm, connect spies}]
    		\node[ rotate=90] at (0,0) {}; outlines={circle,orange,magnification=2,size=3cm, connect spies}]
    		\node[rotate=0, line width=0.05mm, draw=white] at (0,0) { \includegraphics[height=\sz]{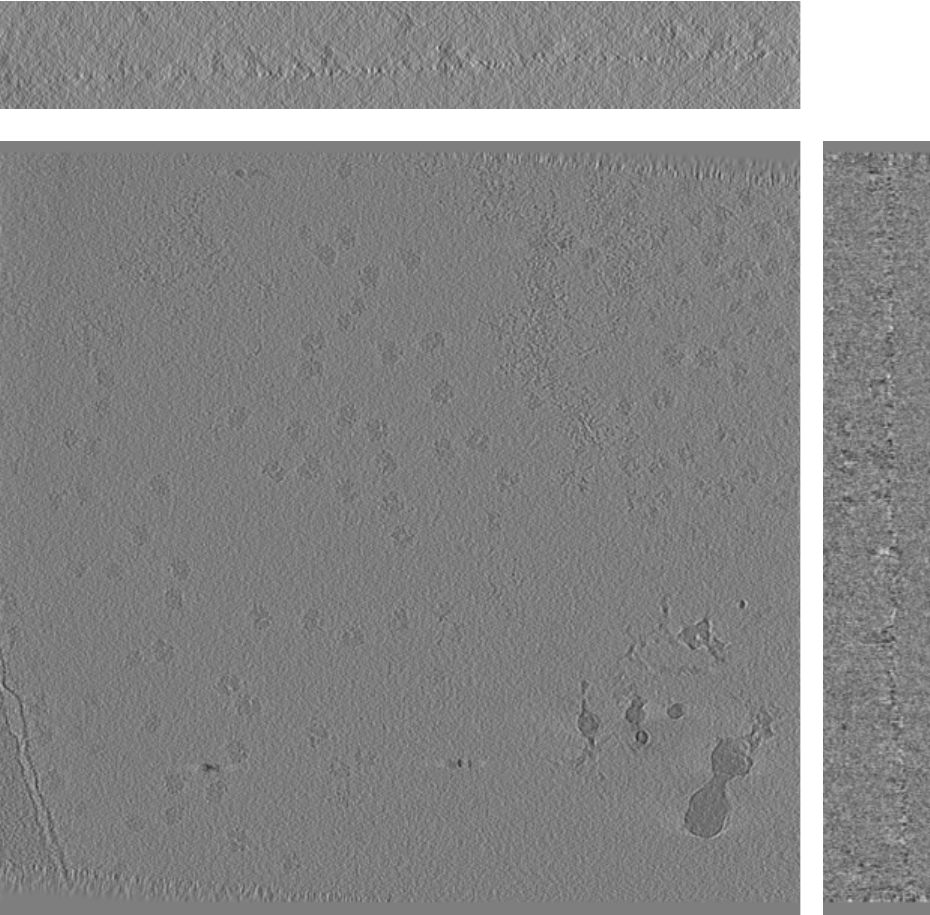}};
    		\spy on (-0.2,0.4) in node [left] at (0.1,-0.8);
		\end{tikzpicture} 
		\caption{FBP ($\sim$ 2 min )}
	\end{subfigure}\hfill
        \begin{subfigure}[t]{\ps\textwidth}
	    \centering
    	\begin{tikzpicture}[spy using outlines={circle,orange,magnification=4,size=2cm, connect spies}]
    		\node[ rotate=90] at (0,0) {}; outlines={circle,orange,magnification=2,size=3cm, connect spies}]
    		\node[rotate=0, line width=0.05mm, draw=white] at (0,0) { \includegraphics[height=\sz]{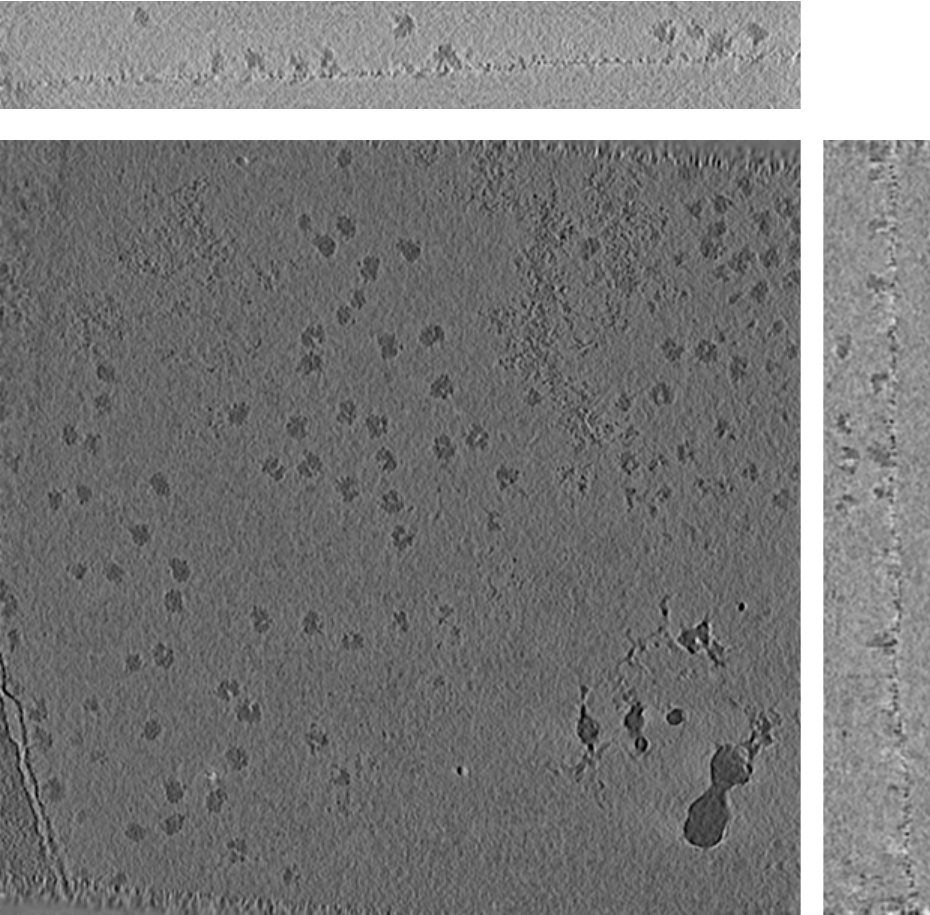}};
    		\spy on (-0.2,0.4) in node [left] at (0.1,-0.8);
		\end{tikzpicture} 
		\caption{CryoCARE+IsoNet ($\sim$ 1 day )}
	\end{subfigure}\hfill
        \begin{subfigure}[t]{\ps\textwidth}
	    \centering
    	\begin{tikzpicture}[spy using outlines={circle,orange,magnification=4,size=2cm, connect spies}]
    		\node[ rotate=90] at (0,0) {}; outlines={circle,orange,magnification=2,size=3cm, connect spies}]
    		\node[rotate=0, line width=0.05mm, draw=white] at (0,0) { \includegraphics[height=\sz]{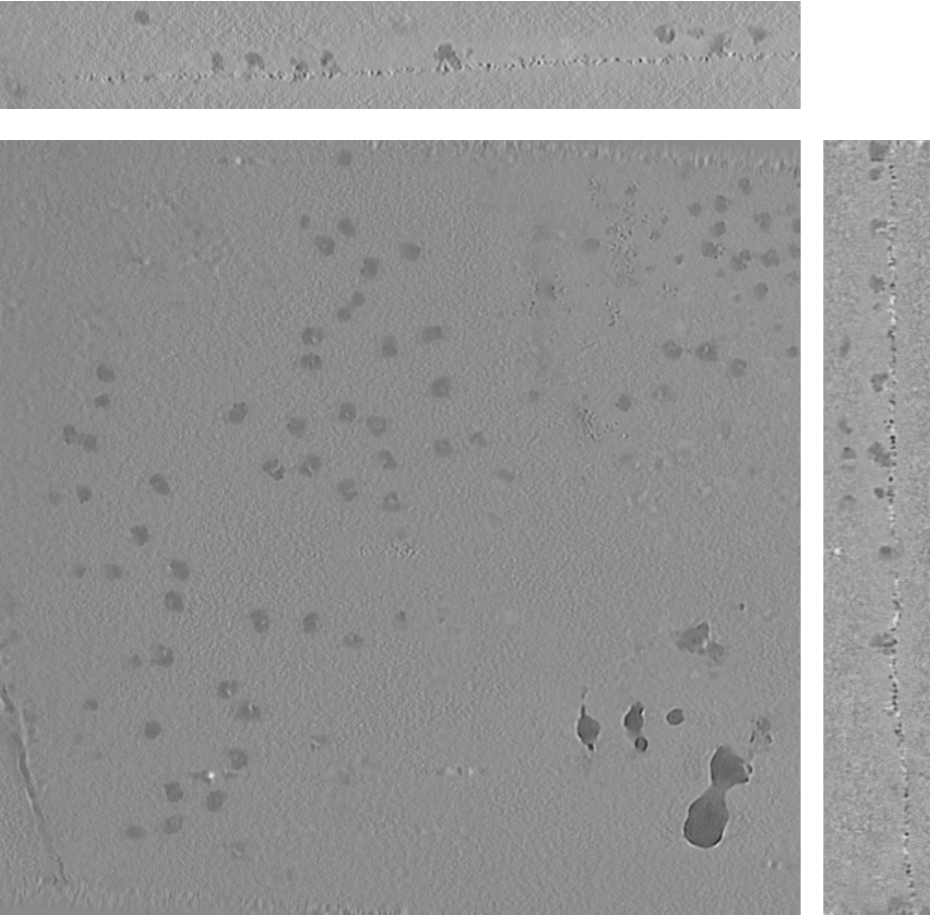}};
    		\spy on (-0.2,0.4) in node [left] at (0.1,-0.8);
		\end{tikzpicture} 
		\caption{DeepDeWedge ($\sim$ 1 day )}
	\end{subfigure}\hfill
	\begin{subfigure}[t]{\ps\textwidth}
	    \centering
    	\begin{tikzpicture}[spy using outlines={circle,orange,magnification=4,size=2cm, connect spies}]
    		\node[ rotate=90] at (0,0) {}; outlines={circle,orange,magnification=2,size=3cm, connect spies}]
    		\node[rotate=0, line width=0.05mm, draw=white] at (0,0) { \includegraphics[height=\sz]{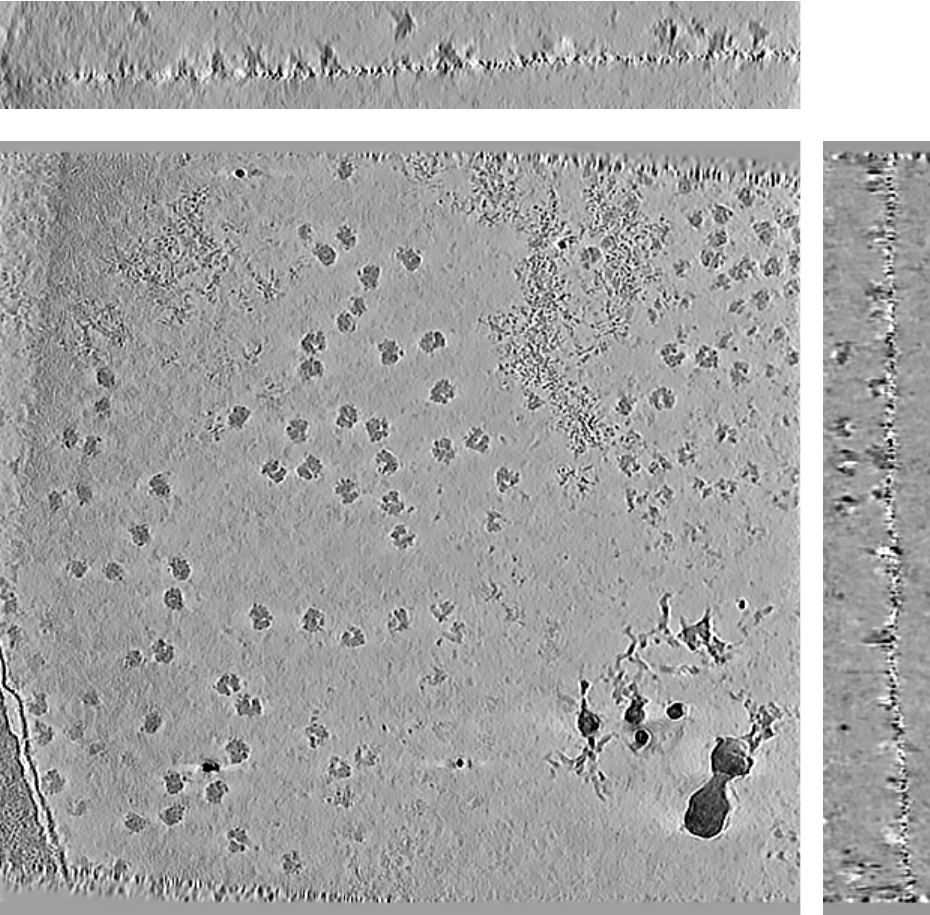}};
    		\spy on (-0.2,0.4) in node [left] at (0.1,-0.8);
		\end{tikzpicture} 
		\caption{Ours ($\sim$ 40 mins )}
	\end{subfigure}
\caption{Evaluation of the model trained on real volumes measurement pairs, using the 80s ribosome projections.}\label{fig:real-recons-ribo}
\end{figure*}

\section{Introduction}
Cryogenic Electron tomography (cryo-ET) is an imaging technique which enables visualization of 3D cellular samples at nanometer resolution, preserving the native context of cellular structures, thus providing insights into their biological process \cite{navarro2022quantitative}. The samples are plunge frozen to obtain samples on the amorphous ice and images using an electron microscope. The measurements are acquired by tilting the frozen samples to a predefined set of angles. Since the measurements are tilted to a limited set of angles, we do not acquire the complete information of the object. This is characterized by a missing wedge of frequencies in the Fourier domain which makes the reconstruction problem is ill-posed. The forward model can be described as 

\begin{equation}
    y^*_{\theta}(r_x,r_y) = \int V(\vR_{\theta}\vr)dr_z, \vr = (r_x,r_y,r_z), 
\end{equation}
where $V(\cdot): \mathbb{R}^{3} \mapsto \mathbb{R}$ is the volume density of the sample and $\vR_{\theta}$ is the 3D rotation matrix along a fixed axis. The tilts are typically between $-60^{\circ}$ and $60^{\circ}$, leading to the missing wedge problem. Further, due to the low electron density while measuring the volume, we observe noisy projections. Let $\mathcal{P}(\cdot)$ be the random noise operator which models the noise in the observation, then: 

\begin{equation*}
    y_{\theta}(r_x,r_y) \sim \mathcal{P}(y^*_{\theta}(r_x,r_y))
\end{equation*}

\subsection{Related Work}
Filtered Back projection (FBP) obtained from the noisy measurements appears historically as the procedure to denoise and reconstruct the underlying cryo-ET volumes. Recently, self-supervised methods were popularized to compensate for both noise and the missing wedge.  Cryo-CARE\cite{buchholz2019cryo} is a popular method in cry-ET to denoise the reconstructions, where a 3D UNet\cite{ronneberger2015u} is trained in a Noise2Noise\cite{lehtinen2018noise2noise} manner.
Cryo-CARE relies on the setting where multiple frames are recorded at each tilt and grouped into two similar sets of projections with different noise realizations. Cryo-CARE first recovers two reconstructions using FBP and then trains the denoising model.
To compensate for the missing wedge, IsoNet \cite{liu2022isotropic} has gained popularity. This self-supervised method trains a 3D UNet on sub-parts of FBP reconstruction. IsoNet artificially removes the wedge of frequencies in the Fourier domain and uses a network to predict it. At evaluation time, the network is used to estimate the frequencies at missing locations. Finally, DeepDeWedge\cite{wiedemann2024deep} has been proposed  as an extension of IsoNet to incorporate both denoising and missing wedge filling jointly using a single Network.

Over the years, self-supervised have been preferred to supervised ones, as it is almost impossible to obtain reference data in cryoET.
However, these approaches require large training time for each new measurement and careful tuning of the hyperparameters, increasing the training time. The reliance on initial 3D reconstruction and large 3D UNets further increases the computational and memory footprint. Recently,  khorashadizadeh \text{et. al.} \cite{khorashadizadeh2025glimpse} introduced a local reconstruction network using only small Multi-Layer Perceptrons (MLP) to perform reconstruction on 2D Computed Tomography (CT). They showed that the network is memory efficient and robust to distribution shifts. In this work, we extend this idea in the setting of 3D cryo-ET. We show that we can extract a localized region in the measurement domain and recover the volumes in a point-wise manner with the help of MLP.  We train these networks in a supervised manner using reconstructions obtained from self-supervised methods. 




\section{Method}
The standard FBP method involves filtering the projection with a ramp filter and then applying back-projection to obtain the volumes. We let $\tilde{y}_{\theta}$ denote the ramp filtered projection, then the back-projected volume at the location $\vr = (r_x,r_y,r_z)$ is obtained as 

\begin{equation}
    \hat{V}_{\text{FBP}}(\vr) = \int_{\theta} \tilde{y}_{\theta}(r_x\cos(\theta) - r_z\sin(\theta) , r_y)d\theta = \int_{\theta} \tilde{y}_{\theta}(\vr_{\theta})d\theta
\end{equation}
The back projection operator provides the locations of the filtered projections that need to be sampled to recover the volumes at a location $\vr$.  However, the FBP can recover the volume perfectly, providing the full range of measurements under no-noise conditions. To recover the volume from finite, limited, and noisy measurements, we propose to extract a fixed-size patch around the locations used in the back projection operator. Further, we replace the averaging operation (the integral), by a more expressive aggregation operator using an MLP. More precisely, to estimate the volume at location $\vr$, we extract patches from the filtered projections centered at location $\vr_{\theta}$. Let $C(\cdot, \cdot)$ be this patch extraction operator, and Let $P\times P$ denote the size of the patch that has to be extracted. Then:
\begin{equation}
    C(\vy_{\theta},\vr)[i,j] =\vy_{\theta} \bigg(r_{\theta}(\vr)- \Delta \begin{bmatrix}
        i \\ j 
    \end{bmatrix}\bigg) , i,j \in \bigg(-\bigg\lfloor \frac{P}{2} \bigg\rfloor, \bigg\lfloor \frac{P}{2} \bigg\rfloor \bigg),
\end{equation}
where $\Delta$ is a scalar, which depends on the resolution of the projection.

Provided the reference volume, the network (MLP) $f$ is trained in a supervised manner by solving the following optimization problem
\begin{equation}\label{eq:optim_super}
    \min_{f}\mathbb{E}_{V,y,\vr}\| V(\vr)  - f(\vp)\|^2,
\end{equation}
where 
\begin{equation}\label{eq:optim_super}
   \vp = [C(\vy_{\theta_1}, r_{\theta_1}(\vr)), \dots, C(\vy_{\theta_N},  r_{\theta_N}(\vr)) ].
\end{equation}

\section{Experiments and Results}
Since Cryo-ET doesn't have any reference volumes or volumes that have been recovered without any artifacts, we rely on volumes recovered from state-of-the-art self-supervised methods. Particularly, we use reconstructions from Cryo-CARE followed by IsoNet (Cryo-CARE + IsoNet) as the reference. We choose a subset of the data from EMPIAR 11830 \cite{kelley2024towards} dataset. These contain Chlamydomonas reinhardtii samples prepared using cryo-plasma Fib milling. The dataset contains projections of size  $4096 \times 4096$ measured at resolution 1.96\AA. We chose the tilt series that contains 41 projections and manually inspected the Cryo-CARE reconstruction provided by the authors and discarded those that looked visually noisy. We then used IsoNet to fill the missing wedge and used these reconstructions as reference volumes. Note that the projections were $4\times$ downsampled first to obtain the volumes. We chose 13 volumes, of which 10 were used for training and the reaming for testing and validation.

We compare the reconstruction of the network on the test volume, visually and using the Fourier Shell Correlation Metric (FSC) \cite{harauz1986exact} a popular metric used to evaluate reconstruction in Cryo-ET.  Figure \ref{fig:real-test} shows the orthogonal slices of the reconstruction on the test set along with the FSC curve. We observe that visually, our network recovers the volumes similar to the reference data. This is confirmed by the FSC metric, where the FSC for our approach is close to one in all frequencies.

To test the model capabilities on a general set of measurements, we consider EMPIAR-10045 \cite{bharat2016resolving} dataset. This contains 7 tilts series of purified S. cerevisiae 80S Ribosomes. The tilt series contains 41 aligned projections between -60 to 60 degrees at  a resolution of $2.17$\AA. We use the tilt series from 'tomogram 5'. Further, we downsample the tilt series by a factor of 4.  Figure \ref{fig:real-recons-ribo} shows the orthogonal slices of the reconstruction for FBP, Cryo-CARE+IsoNet, DeepDeWedge, and ours. We observe that our approach recovers the denoised volumes for the new dataset in a fraction of the time required for self-supervised methods. More detailed experiments on synthetic and real datasets  are provided in \cite{kishore2025end}.

\section{Conclusion}
We present a supervised method for Cryo-ET reconstructions. Using the localized patches, we can recover the volumes directly from noisy projections with limited angles similar to or better than the current methods used in practice. Further, due to the point-wise nature of the network, it can be trained with a few sets of measurements and still recovers volumes different from the ones used in training. Along with this paper, we provide a pre-trained network that can be used to avoid the long training time usually present in self-supervised methods. As the network recovers the volume one voxel at a time, there is a flexible trade between reconstruction time and complexity without any loss in the quality of reconstruction.

\bibliographystyle{IEEEtran}
\bibliography{refs}

\begin{thebibliography}{10}
\providecommand{\url}[1]{#1}
\csname url@samestyle\endcsname
\providecommand{\newblock}{\relax}
\providecommand{\bibinfo}[2]{#2}
\providecommand{\BIBentrySTDinterwordspacing}{\spaceskip=0pt\relax}
\providecommand{\BIBentryALTinterwordstretchfactor}{4}
\providecommand{\BIBentryALTinterwordspacing}{\spaceskip=\fontdimen2\font plus
\BIBentryALTinterwordstretchfactor\fontdimen3\font minus \fontdimen4\font\relax}
\providecommand{\BIBforeignlanguage}[2]{{%
\expandafter\ifx\csname l@#1\endcsname\relax
\typeout{** WARNING: IEEEtran.bst: No hyphenation pattern has been}%
\typeout{** loaded for the language `#1'. Using the pattern for}%
\typeout{** the default language instead.}%
\else
\language=\csname l@#1\endcsname
\fi
#2}}
\providecommand{\BIBdecl}{\relax}
\BIBdecl

\bibitem{navarro2022quantitative}
P.~P. Navarro, ``Quantitative cryo-electron tomography,'' \emph{Frontiers in Molecular Biosciences}, vol.~9, p. 934465, 2022.

\bibitem{buchholz2019cryo}
T.-O. Buchholz, M.~Jordan, G.~Pigino, and F.~Jug, ``Cryo-{CARE}: Content-aware image restoration for cryo-transmission electron microscopy data,'' in \emph{2019 IEEE 16th International Symposium on Biomedical Imaging (ISBI 2019)}, IEEE, Ed.\hskip 1em plus 0.5em minus 0.4em\relax IEEE, 2019, pp. 502--506.

\bibitem{ronneberger2015u}
O.~Ronneberger, P.~Fischer, and T.~Brox, ``U-net: Convolutional networks for biomedical image segmentation,'' in \emph{Medical image computing and computer-assisted intervention--MICCAI 2015: 18th international conference, Munich, Germany, October 5-9, 2015, proceedings, part III 18}, Springer, Ed.\hskip 1em plus 0.5em minus 0.4em\relax Springer, 2015, pp. 234--241.

\bibitem{lehtinen2018noise2noise}
J.~Lehtinen, J.~Munkberg, J.~Hasselgren, S.~Laine, T.~Karras, M.~Aittala, and T.~Aila, ``Noise2noise: Learning image restoration without clean data,'' \emph{arXiv preprint arXiv:1803.04189}, 2018.

\bibitem{liu2022isotropic}
Y.-T. Liu, H.~Zhang, H.~Wang, C.-L. Tao, G.-Q. Bi, and Z.~H. Zhou, ``Isotropic reconstruction for electron tomography with deep learning,'' \emph{Nature communications}, vol.~13, no.~1, p. 6482, 2022.

\bibitem{wiedemann2024deep}
S.~Wiedemann and R.~Heckel, ``A deep learning method for simultaneous denoising and missing wedge reconstruction in cryogenic electron tomography,'' \emph{Nature Communications}, vol.~15, no.~1, p. 8255, 2024.

\bibitem{khorashadizadeh2025glimpse}
A.~Khorashadizadeh, V.~Debarnot, T.~Liu, and I.~Dokmani{\'c}, ``Glimpse: Generalized locality for scalable and robust ct,'' \emph{IEEE Transactions on Medical Imaging}, 2025.

\bibitem{kelley2024towards}
R.~Kelley, S.~Khavnekar, R.~D. Righetto, J.~Heebner, M.~Obr, X.~Zhang, S.~Chakraborty, G.~Tagiltsev, A.~K. Michael, S.~van Dorst \emph{et~al.}, ``Towards community-driven visual proteomics with large-scale cryo-electron tomography of chlamydomonas reinhardtii,'' \emph{BioRxiv}, pp. 2024--12, 2024.

\bibitem{harauz1986exact}
G.~Harauz and M.~van Heel, ``Exact filters for general geometry three dimensional reconstruction.'' \emph{Optik.}, vol.~73, no.~4, pp. 146--156, 1986.

\bibitem{bharat2016resolving}
T.~A. Bharat and S.~H. Scheres, ``Resolving macromolecular structures from electron cryo-tomography data using subtomogram averaging in relion,'' \emph{Nature protocols}, vol.~11, no.~11, pp. 2054--2065, 2016.

\bibitem{kishore2025end}
V.~Kishore, V.~Debarnot, R.~D. Righetto, A.~Khorashadizadeh, B.~D. Engel, and I.~Dokmani{\'c}, ``End-to-end localized deep learning for cryo-et,'' \emph{arXiv preprint arXiv:2501.15246}, 2025.

\end{thebibliography}
\end{document}